# Explanation-Based Auditing

Daniel Fabbri[*] and Kristen LeFevre[*]

{*dfabbri, klefevre*}*@umich.edu*
*Electrical Engineering & Computer Science, University of Michigan, 2260 Hayward Ave. Ann Arbor, MI 48109 USA*

## ABSTRACT

To comply with emerging privacy laws and regulations, it has become common for applications like electronic health records systems (EHRs) to collect *access logs*, which record each time a user (e.g., a hospital employee) accesses a piece of sensitive data (e.g., a patient record). Using the access log, it is easy to answer simple queries (e.g., Who accessed Alice's medical record?), but this often does not provide enough information. In addition to learning who accessed their medical records, patients will likely want to understand *why* each access occurred.

In this paper, we introduce the problem of generating *explanations* for individual records in an access log. The problem is motivated by user-centric auditing applications, and it also provides a novel approach to misuse detection. We develop a framework for modeling explanations which is based on a fundamental observation: For certain classes of databases, including EHRs, the reason for most data accesses can be inferred from data stored elsewhere in the database. For example, if Alice has an appointment with Dr. Dave, this information is stored in the database, and it explains why Dr. Dave looked at Alice's record. Large numbers of data accesses can be explained using general forms called *explanation templates*.

Rather than requiring an administrator to manually specify explanation templates, we propose a set of algorithms for automatically discovering frequent templates from the database (i.e., those that explain a large number of accesses). We also propose techniques for inferring collaborative user groups, which can be used to enhance the quality of the discovered explanations. Finally, we have evaluated our proposed techniques using an access log and data from the University of Michigan Health System. Our results demonstrate that in practice we can provide explanations for over 94% of data accesses in the log.

## 1. INTRODUCTION

In recent years, laws and regulations have imposed a number of new requirements governing the responsible management of personal and private data. For example, in the United States, the

[*]This work was supported by National Science Foundation grants CNS-0915782 and IGERT-0903629.

| Lid | Date | User | Patient |
|---|---|---|---|
| L100 | Mon Jan 03 10:16:57 2010 | Nurse Nick | Alice |
| L116 | Mon Jan 03 11:22:43 2010 | Dr. Dave | Alice |
| L127 | Mon Jan 03 17:09:03 2010 | Radiologist Ron | Alice |
| L900 | Mon Apr 28 14:29:08 2010 | Surgeon Sam | Alice |

(Annotation: Alice had an appointment with Dr. Dave on Jan 3, 2010. → L116)

**Figure 1: Sample access log and explanation**

Health Insurance Portability and Accountability Act (HIPAA) stipulates that individuals have the right to request an accounting of the disclosures of their protected health information (PHI) by hospitals and other healthcare providers (so-called "covered entities"). Recently, the U.S. Department of Health and Human Services proposed an expansion of this rule, which would require covered entities to provide individuals with detailed *access reports*, including the names of all people who have accessed their electronic PHI.[1]

Most modern electronic health records systems (EHRs) collect *access logs* automatically. For example, the University of Michigan Health System has built and deployed a web-based clinical EHR system called CareWeb[2]. To support regulatory compliance, each time an employee accesses a medical record via CareWeb, a record is added to the access log. While the precise format can vary among EHR systems, it is typically quite simple. CareWeb access logs contain four main attributes: *Timestamp*, *User_ID*, *Patient_ID*, and a coded description of the *Action* performed (e.g., *viewed lab reports*, or *updated history*).

One promising approach to providing access reports, and improving overall transparency, is the idea of *user-centric auditing*. Basically, the idea is to construct a portal where individual patients can login and view a list of all accesses to their medical records. When the underlying access logs are of the form described above, this is relatively straightforward. Unfortunately, the resulting access histories are often long and hard to analyze. Worse, the list of accesses often includes accesses by many people the patient does not know. (For example, the patient probably knows the name of his primary care physician, but he is not likely to recognize the name of the intake nurse or the radiologist who read his x-ray.)

In this paper, we observe that in addition to asking who has accessed their medical records, patients will want to understand *why* these people accessed their records.

EXAMPLE 1.1. *Consider a patient Alice who is using a user-centric auditing system. She logs into the patient portal and requests a log of all accesses to her medical record. The resulting log is shown in Figure 1, and includes accesses by four different hospital employees.*

---
[1]HHS Press Release, May 31, 2011.
http://www.hhs.gov/news/press/2011pres/05/20110531c.html
[2]http://www.med.umich.edu/mcit/carewebwe/help/overview.html

Looking at this log, Alice would like to understand the reason for each of these accesses. Ideally, we would like to provide an explanation for each access; if Alice clicks on a log record, she should be presented with a short snippet of text:

- **L100** Nurse Nick works with Dr. Dave, and Alice had an appointment with Dr. Dave.
- **L116** Alice had an appointment with Dr. Dave.
- **L127** Radiologist Ron reviewed Alice's x-rays for Dr. Dave.
- **L900** Surgeon Sam performed a surgery for Alice after Dr. Dave referred Alice to Sam.

One approach to providing explanations would require the user (e.g., the doctor) to enter a reason each time he accesses a medical record. While some systems may require this (e.g., [3]), it places a large burden on users.

Another approach would identify the access control rule(s) that allowed access to the medical record. (For example, users with a clinical appointment may be granted access to patient records.) Unfortunately, in environments like hospitals, it is very difficult to specify and maintain detailed access control policies [2]. (For example, residents and medical students change departments as often as once per week.) Further, overly restrictive policies can have disastrous consequences, interfering with patient care. As a result, it is typical for many more users to be granted access to a particular medical record than have a legitimate clinical or operational reason for accessing the record.

Instead, we would like to develop a technique to automatically produce informative explanations. Of course, there may be accesses for which we are not able to generate explanations. In these cases, if the access appears suspicious, the patient has the right to report the access to the hospital compliance office, and to request an investigation. However, developing a system to generate explanations automatically is useful both for the purpose of informing patients how their medical records are being used and for reducing the burden on the compliance office in handling complaints.

Interestingly, this also suggests a secondary application of explanations for the purpose of automated *misuse detection*. Because of the difficulties in expressing and maintaining access control policies up-front, rather than preventing data access, hospitals often err on the side of maintaining an access log in hopes of detecting misuse after the fact. Unfortunately, there are few technical tools for proactively detecting misuse from the access log. Common approaches often involve manual analysis in response to a complaint, or monitoring accesses to the medical records of VIPs (high-profile people).[3] Of course, manual analysis does not scale to the access logs collected by modern hospitals. (For example, in just one week, the University of Michigan Health System collected over 4 million access log records via CareWeb.) On the other hand, if we are able to automatically construct explanations for why accesses occurred, we can conceivably use this information to reduce the set of accesses that must be examined to those that are unexplained. While we are not likely to be able to explain every access, this process significantly reduces the set of records that are potentially suspicious.

## 1.1 Contributions

In this paper, we study the novel problem of automatically explaining individual log records (accesses) in an access log. Our work is inspired by a fundamental observation: *For certain classes of databases, including those used to store EHR data, there is typically a clear reason for each access. Further, this reason can often be gleaned from information stored elsewhere in the database.* We provide an extensive empirical study in Section 5 using a large access log and EHR data from the Michigan Health System (CareWeb), which validates our hypothesis. Based in part on this observation, we make the following important contributions:

- In Section 2 we define a novel approach to modeling explanations. Intuitively, an explanation can be viewed as a connection from the data accessed (e.g., the Patient), through the database, and back to the user who accessed the data (e.g., the User).
- Before explanations can be used, they must be generated or specified. Our empirical study indicates that most accesses can actually be explained using a limited number of explanation types, or *templates*. For example, the fact that a patient had an appointment with the user who accessed his record is a general explanation type that can explain many different accesses.
- Nonetheless, we would like to remove some of the burden from the administrator in specifying explanation templates. Thus, in Section 3 we propose algorithms for automatically discovering templates that occur frequently in a given database (i.e., that explain a large number of accesses).
- We observe that databases such as CareWeb are often missing information that is useful for the purpose of constructing explanations. For example, Dr. Dave and Nurse Nick work together, but this information is not recorded anywhere. In Section 4 we describe techniques to infer some of this missing data so that more accesses can be explained.
- Finally, in Section 5, we describe an extensive empirical study and experimental evaluation using data from CareWeb, which contains over 4.5 million accesses as well as records of appointments, visits, documents produced, and other information. Our experiments confirm the hypothesis that there is a reason for most accesses in our log, and that these accesses can be explained using data located elsewhere in the database. Further, the experiments indicate that (i) common explanation templates can be mined automatically, (ii) missing data can be added to the database to improve the rate at which accesses are explained without producing a significant number of false positives, and (iii) the explanations discovered can explain over 94% of accesses in the log.

## 2. EXPLAINING ACCESSES

Given an entry in an access log, which describes both the data that was accessed (e.g., the patient's medical record) and the user who accessed the data, our goal is to construct a simple *explanation* describing the reason for the access. In addition, an explanation should satisfy the following basic properties:

- **Human Interpretable:** The reason why the access occurred should be easily understood. Among other things, we argue that an explanation should be logical and boolean (either it explains the access or not). In contrast, systems that provide probability distributions or other ambiguity are difficult to interpret.
- **General:** Explanations should take on a general form whereby a single explanation type or *template* explains many accesses by many users. For example, a patient having an appointment with the doctor who accesses his medical record is a common explanation template that can be used to explain many different accesses in the log.
- **Concise:** The explanation should be represented concisely.
- **Easy to produce/calculate:** Given a particular access, it should be easy to compute the explanation(s) for the access.

---
[3]For example, in 2008, hospital employees inappropriately accessed Britney Spears' medical record [22]. Also, in 2008, U.S. State Department employees were fired for inappropriately accessing President Obama's passport file [16].

## 2.1 Explanation Templates

We begin by formalizing the structure of explanations, which can be used to describe why individual accesses occurred. We model an explanation based on the hypothesis that for every legitimate data access, there is a reason for the access, and in most cases the reason can be gleaned from information stored elsewhere in the database.

EXAMPLE 2.1. *Consider the patients Alice and Bob, who log into the patient portal of their healthcare provider's electronic medical records system. To support transparency, the portal allows the patients to view a log of hospital employees who have accessed their records. Among others, the patients observe that an employee named Dr. Dave accessed their medical records. While this information may itself be useful, oftentimes it is important to provide further details, explaining why Dr. Dave accessed the record. Consider the following possible explanations:*

A. *Dr. Dave accessed Alice's medical record because Alice had an appointment with Dr. Dave on 1/1/2010.*

B. *Dr. Dave accessed Bob's medical record because Bob had an appointment with Dr. Mike on 2/2/10, and Dr. Dave and Dr. Mike work together in the Pediatrics department.*

C. *Dr. Dave accessed Alice's medical record because Dr. Dave previously accessed Alice's record.*

D. *Dr. Dave accessed Alice's medical record because Alice had an appointment with someone else.*

Intuitively, an explanation should connect the user who accessed the data with the data itself (i.e., the patient's medical record). In examples (A-C), notice that there is a connection from the user who accessed the data (Dr. Dave), through the data in the database (appointment and department information), back to the data that was accessed (Alice or Bob's medical record). In contrast, the final explanation does not provide a connection between the user and the data. Consequently, the final explanation does not provide a meaningful description of why Dr. Dave in particular accessed Alice's record.

To capture this intuition more formally, we can model the explanation as a *path* through the database, beginning and ending in the log. We assume that the database stores a log of accesses, which records the time of the access, the user who accessed the data (Log.User) and a reference to the data that was accessed (Log.Patient). An *explanation template* is a tool that can be used to explain many individual accesses.

DEFINITION 1 (EXPLANATION TEMPLATE). *An explanation template is a stylized query on the database and log.*

*Consider a query Q of the following form, where $T_1, ..., T_n$ are (not necessarily distinct) tables in the database, and each $C_i$ is an attribute comparison condition of the form $A_1 \theta A_2$ where $\theta \in \{<, \leq, =, \geq, >\}$.*

```
SELECT Log.Lid, A_1, ..., A_m
FROM Log, T_1, ..., T_n
WHERE C_1 AND ... AND C_j
```

*Let G be a graph, where each attribute in $Log, T_1, ..., T_n$ is a node. Let there be an edge from attribute $A_1$ to $A_2$ in G if (i) $A_1$ and $A_2$ are in the same tuple variable (i.e., $Log, T_1, ..., T_n$) or (ii) Q imposes a comparison condition between $A_1$ and $A_2$.*

*Query Q is an explanation template if there is a* path *P on G that starts at the data that was accessed (Log.Patient) and terminates at the user who accessed the data (Log.User), touching at least one attribute from each tuple variable mentioned in the query, and where no edge is traversed more than once.*

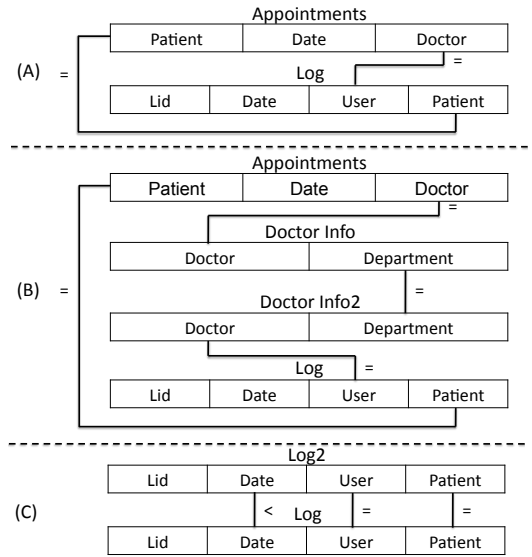

Figure 2: Paths through the explanation graph

| Patient | Date | Doctor |
|---|---|---|
| Alice | 1/1/2010 | Dave |
| Bob | 2/2/2010 | Mike |

(a) Appointments

| Doctor | Dept. |
|---|---|
| Mike | Pediatrics |
| Dave | Pediatrics |

(b) Doctor Info

| Lid | Date | User | Patient |
|---|---|---|---|
| L1 | 1/1/2010 | Dave | Alice |
| L2 | 2/2/2010 | Dave | Bob |

(c) Log

Figure 3: Example hospital database and log of accesses

Because an explanation template is a query, it can be used to explain why many different data accesses occurred. We refer to these data-specific descriptions (query results) as *explanation instances*. Notice that instances of a particular explanation template can be easily converted to natural language by providing a parameterized *description string*.

EXAMPLE 2.2. *Consider the database and log provided in Figure 3. Explanations like (A) from Example 2.1 can be derived from the following explanation template:*

```
SELECT L.Lid, L.Patient, L.User, A.Date
FROM Log L, Appointments A
WHERE L.Patient = A.Patient
   AND A.Doctor = L.User
```

*Figure 2 (A) shows the graph G associated with this explanation template. Notice that there is a path P that starts at $Log.Patient$ and terminates at $Log.User$. The edges between attributes in the same tuple variable are implicit.*

*Instances of this explanation template can easily be converted to natural language using a simple description string: "[L.Patient] had an appointment with [L.User] on [A.Date]." For example, log record L1 can be explained using the description "Alice had an appointment with Dave on 1/1/2010."*

*Explanations like example (B) can be derived from the following explanation template:*

```
SELECT L.Lid, L.Patient, L.User, A.Doctor,
   A.Date, I1.Department
FROM Log L, Appointments A, Doctor_Info I1,
```

```
    Doctor_Info I2
WHERE L.Patient = A.Patient
   AND A.Doctor = I1.Doctor
   AND I1.Department = I2.Department
   AND I2.Doctor = L.User
```

*Figure 2 (B) shows the graph associated with this explanation template. Instances of this explanation are easily expressed in natural language: "[L.Patient] had an appointment with [A.Doctor] on [A.Date], and [L.User] and [A.Doctor] work together in the [I1.Department] department."*

Notice that a single log record may have multiple explanation instances, generated from one or more explanation templates. For example, the query implementing explanation (A) would produce multiple results with Lid = L1 if Alice had multiple appointments with Dr. Dave. We consider each of these instances to be a valuable source of information; in practice, when there are multiple explanation instances for a given log record, we convert each to natural language and rank the explanations in ascending order of path length.

It is useful to draw a further distinction between what we will call *simple* explanation templates and their more complex *decorated* counterparts.

DEFINITION 2 (SIMPLE EXPLANATION TEMPLATE). *Consider an explanation template and its associated graph $G$ and path $P$. The explanation template is* simple *if it is not possible to remove any set of selection condition edges from $G$ and still have a path $P'$ from Log.Patient to Log.User.*

Intuitively, a simple explanation provides a minimal connection between the data and the user who accessed it. Notice that explanation templates (A) and (B) in Example 2.2 are both simple.

At the same time, simple explanations may not always be sufficient to express the desired semantics. For example, suppose we want to express the idea that an access occurred because the same user previously accessed the data (e.g., explanation (C) in Example 2.1). A simple explanation template could partially capture the desired semantics as follows:

```
SELECT L1.Lid, L1.Patient, L1.User
FROM Log L1, Log L2
WHERE L1.Patient = L2.Patient
   AND L2.User = L1.User
```

However, to express the temporal aspect of the explanation, we need the additional selection condition `L1.Date > L2.Date`. Figure 2 (C) shows the graph associated with this explanation template. As a result, this *decorated* explanation always explains a subset of the accesses that are explained by the corresponding simple explanation.

DEFINITION 3 (DECORATED EXPLANATION TEMPLATE). *A decorated explanation template is a simple explanation template with additional selection conditions added.*

Finally, for modern databases with large schemas, the number and complexity of explanations can be very large, even if we only consider simple explanations. At the same time, we hypothesize that most explanations only require information from a few tables in the database. (We verify this hypothesis in Section 5.) For this reason, we may restrict the number of tables that a path can reference to an administrator-specified value $T$.

DEFINITION 4 (RESTRICTED EXPLANATION TEMPLATE). *A restricted simple explanation template is a simple explanation template that only refers to at most $T$ tables.*

## 3. MINING EXPLANATIONS

Before explanations can be used in any particular database, the appropriate explanation templates must be specified. One approach would require the security or database administrator to specify explanation templates manually. However, this can be a tedious process. Worse, due to the complexity of modern database schemas, a single administrator may not have complete knowledge of all the different reasons that data accesses occur.

While it is important to keep the administrator in the loop, we argue that the system should reduce the administrator's burden by automatically suggesting templates from the data. In this section, we describe our approach to mining templates from a given database. The administrator can then review the suggested set of templates before applying them.

The goal of the mining algorithms is to find the set of *frequent* explanation templates, or those that can be used to explain many accesses. Intuitively, this reduces the possibility of spurious results. The problem of mining frequent explanation templates is related to previous work on frequent pattern mining [5]. Indeed, our algorithms take a bottom-up pruning approach inspired by algorithms like *a priori*. At the same time, there are several important differences between the template mining problem and frequent pattern mining that prevent us from directly applying existing algorithms: First, we are mining connected paths between a start and end attribute in the schema. Second, our measure of frequency (support) is different; for explanation templates, frequency is determined by the number of accesses in the log that are explained by the template, so every path we consider must reference the log. Finally, the data is stored across multiple tables in the database, rather than in a single large file of transactions.

### 3.1 Problem Statement

Our goal is to find the set of explanation templates that occur frequently in a given database instance. We define $support$ to be the number of accesses in the log that are explained by the template.

An extremely naive approach would enumerate all possible templates of the form described in Definition 1. However, the number of possible templates is unbounded. Even if we restrict ourselves to simple templates without self-joins, the number of possible templates is still exponential in terms of the number of attributes in the schema.

To reduce the space, we make some practical simplifying assumptions: (1) We only consider simple explanation templates. (2) We only consider equi-joins between two tables if there exists a key-foreign key relationship, or if another relationship between two attributes is explicitly provided by the administrator. (3) An attribute and table can only be used in a self-join if the administrator explicitly allows the attribute to be used in a self-join. (4) We restrict the path length to $M$ and restrict the number of tables referenced to $T$. We leave the task of developing algorithms for mining more complex (decorated) explanation templates to future work.

DEFINITION 5 (EXPLANATION MINING). *Given a database $D$ and a log of accesses $L$, return those explanation templates of length at most $M$, that reference at most $T$ tables and that explain (support) at least $s\%$ of the accesses in the log, where the edges in the path are restricted to attributes from the same tuple variable, key relationships, specified self-joins, or administrator-specified relationships.*

EXAMPLE 3.1. *Continuing with Example 2.1 and the database in Figure 3, template (A) has support of 50% (from access L1), and template (B) has support of 100% (from accesses L1 and L2).*

**Algorithm 1** One-Way Template Mining Algorithm
---
**Input:** Start attribute (Log.Patient), end attribute (Log.User), support ($S$), max path length (M), restricted number of tables referenced ($T$), the set of edges from the schema (Edges) and the database instance ($D$).
**Output:** Set of supported explanation templates (up to the max length).
1: Length = 1
2: Paths = {Edges that begin with the start attribute}
3: Explanations = {}
4: **while** Length $\leq$ M **do**
5:    New Paths = {}
6:    **for** Path $p \in Paths$ **do**
7:      **for** Edge $e \in Edges$ **do**
8:        **if** areConnected(p, e) **then**
9:          Candidate Path = $p.append(e)$
10:          **if** isARestrictedSimplePath(Candidate Path) **then**
11:            **if** Support(Candidate Path, D) $\geq$ S **then**
12:              New Paths.add(Candidate Path)
13:              **if** isAnExplanation(Candidate Path) **then**
14:                Explanations.add(Candidate Path)
15:    Paths = New Paths
16:    Length += 1
17: Return Explanations
---

## 3.2 One-Way Algorithm

We begin by describing a basic algorithm. (Details are provided in Algorithm 1.) The administrator provides the start attribute Log.Patient (the data that is accessed), the end attribute Log.User (the user who accessed the data), the minimum support $S$, the maximum length $M$, the maximum number of tables referenced $T$, the schema, and the database. We restrict the set of edges (denoted as *Edges*) that can be used in explanations as described in Section 3.1. An initial set of paths of length one are created by taking the set of edges that begin with the start attribute Log.Patient. The goal of the algorithm is to find the set of *supported* explanation templates, which are those templates that explain at least $s\%$ of the accesses.

The algorithm finds the set of supported templates as follows: First, for each path at the current length, and for each edge, the algorithm tests if the two are connected. Intuitively, the path and edge are connected if the last attribute in the path is the same attribute as the first attribute in the edge. Second, for those connected paths and edges, the path and edge are combined by appending the edge to the right end of the path. Third, the algorithm checks if this *candidate* path is a restricted simple path. Intuitively, the candidate path is simple if it begins at the log and continues to join with previously untraversed tables until the log is reached (the path traverses each node at most once and at most two nodes per table). The candidate path is a restricted simple path if it references no more than $T$ tables (a path that references a table and a self-join for that table is counted as a single reference). Next, the candidate path is converted to SQL and evaluated on the database to calculate the path's support. We calculate the support using the following query:

```
SELECT COUNT(DISTINCT Log.Lid)
FROM Log, T_1, ..., T_N
WHERE C
```

If the support is greater than or equal to $S = |Log| \times s\%$, then the path is added to the set of new paths that will be used in the next iteration of the algorithm. Furthermore, if the path has the appropriate start and end attributes, then the path is also an explanation template, and is marked accordingly. The algorithm repeats for paths of increasing length until the maximum path length is reached.

EXAMPLE 3.2. *Consider the database shown in Figure 3; the one-way algorithm works as follows: The input set of edges includes key-foreign key equi-joins such as {Log.Patient = Appointments.Patient, Appointments.Patient = Log.Patient, Log.User = Appointments.Doctor, Appointments.Doctor =Log.User} and the administrator-provided self-join {Doctor_Info.Department = Doctor_Info2.Department}. The initial set of paths is: {Log.Patient = Appointments.Patient}. This first path is converted into SQL and has the selection condition* `Log.Patient = Appointments.Patient` *and is evaluated on the database. The path has support of 100%.*

*Next, connected edges are appended onto the path. For example, one candidate path has the selection condition:* `Log.Patient = Appointments.Patient AND Appointments.Doctor = Log.User`. *This candidate path is also an explanation since it has the correct start and end attributes. The explanation has support of 50%.*

The one-way algorithm works in a bottom-up manner to find the supported explanation templates. We observe several important properties of the algorithm: First, the paths must always include an attribute from the Log in order to calculate the support for the path; if there was no Log attribute, then it would be impossible to count the number of log entries explained.

Second, the support function is monotonic. If a path $P$ of length $\ell - 1$ does not have the necessary support (i.e., does not explain $\geq s\%$ of the accesses in the log), then adding additional edges to the path will never produce an explanation template with the necessary support. Thus, the bottom-up algorithm is able to prune certain paths that are guaranteed not to have the necessary support.

### 3.2.1 Performance Optimizations

We apply three performance optimizations for the algorithm:

**Caching Selection Conditions and Support Values:** We observe that multiple paths may have the same selection conditions, even though the paths traverse the explanation graph in different orders. Since the order in which the selection conditions are applied does not change the result, these paths are guaranteed to have the same support (i.e., $R.attr = T.attr$ is equivalent to $T.attr = R.attr$). Thus, a simple optimization is to cache the support of each path that has already been tested. Then, before the next path's support is calculated, the algorithm checks if some variation of the path (with an equivalent selection condition) has already been tested. If so, the algorithm does not need to evaluate the query on the database and can use the previously recorded support value instead.

**Reducing Result Multiplicity:** The multiplicity of data in the database can impact performance. For example, from Example 2.2, if Alice had three appointments with Dr. Dave, then there would be three instances of explanation (A) for the same log id. These additional rows in the output make computing the support (i.e., the distinct set of log ids) more costly. Therefore, since it does not matter how many times a given log id is in the result, the performance can be improved by reducing the number of rows in the result. To remove duplicates from each table, we use a subquery to extract the distinct set of rows from the table, while only projecting those attributes needed for the path. For example, the query from Example 2.2 can be rewritten as follows:

```
SELECT COUNT(DISTINCT L.Lid)
FROM Log L,
    (SELECT DISTINCT Patient, Doctor
     FROM Appointments) A
WHERE L.Patient = A.Patient
    AND A.Doctor = L.User
```

**Skipping Non-Selective Paths:** For many (short) paths, the selection conditions are not selective and return most of the log. Computing the support for these paths wastes time because these non-selective paths typically have sufficient support and are not pruned. Therefore, the algorithm's performance can be improved by passing these non-selective paths directly to the next iteration of the algorithm, instead of calculating their support. We determine if a path is likely to have sufficient support by asking the database optimizer for the number of log ids it expects to be in the result of the query. If the value is greater than the desired support $S \times c$ (where $c$ is a constant like 10), the system skips this path and adds it to the set of paths to try in the next iteration of the algorithm. In the special case when the path is also an explanation, the path is not skipped. The constant $c$ is used to account for the optimizer's estimation error. Using this optimization, the system trades off pruning some paths in order to not have to calculate the support of the non-selective paths. Even in the worst case when the database optimizer significantly errs with its estimation, the output set of explanation templates does not change because paths are not discarded; rather, they are tested in the next iteration of the algorithm.

## 3.3 Two-Way Algorithm

Intuitively, the two-way algorithm constructs paths in two directions: from the start to the end, and from the end to the start. The two-way algorithm is initiated with the edges that begin with the start attribute and the edges that terminate with the end attribute. The paths that begin with the start attribute are extended to the right with connected edges until the end attribute is reached (i.e., the one-way algorithm), while the paths that terminate with the end attribute are extended to the left with connected edges until the start attribute is reached. Therefore, an optimized algorithm would have them meet in the middle.

### 3.3.1 Bridging Paths

The one-way and two-way algorithms explore all paths that have the desired support. However, the goal of the algorithms is to find supported explanation templates. Therefore, by enforcing the constraint that paths must start and end with particular attributes, we can restrict the set of paths the algorithms must consider. Moreover, since we have paths extending from the start and end attributes, we can combine, or $bridge$, these paths.

Consider the case where the two-way algorithm has executed and produced all supported paths up to length $\ell$. The algorithm can use these paths to easily construct the set of candidate explanation templates up to length $2\ell - 1$ (the candidate templates are a superset of those templates that have the necessary support). These candidate templates can be produced by connecting those paths that begin with the start attribute to those paths that terminate with the end attribute as shown in Figure 4. The remaining paths that do not start or end with one of these attributes can be ignored.

More concretely, candidate templates of length $n$ ($2 \leq \ell < n \leq 2\ell - 1$) can be produced by taking paths of length $\ell$ that begin with the start attribute and connecting them to paths of length $n - \ell + 1$ that terminate with the end attribute. We say the paths are $bridged$ because the algorithm requires that the edges where the two paths are connected (the bridge edge) are equivalent. As a result, the length of the combined path is one less than the sum of the individual path lengths. Once the set of candidate templates is produced, the support for each candidate is tested.

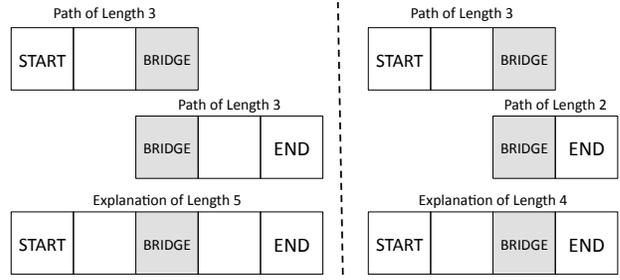

**Figure 4: Bridging paths to create explanations**

EXAMPLE 3.3. *Template (B) from Example 2.1 can be created by bridging the following two paths:*

```
SELECT COUNT(DISTINCT L.Lid)
FROM Log L, Appointments A, Dept_Info I1,
    Dept_Info I2
WHERE L.Patient = A.Patient
    AND A.Doctor = I1.Doctor
    AND I1.Department = I2.Department

SELECT COUNT(DISTINCT L.Lid)
FROM Log L, Dept_Info I1, Dept_Info I2
WHERE I1.Department = I2.Department
    AND I2.Doctor = L.User
```

*Notice that the combined path has the appropriate start and end attributes, and the condition I1.Department = I2.Department can be used to bridge the paths.*

When the length of the desired path is greater than or equal to $2\ell$, the candidates cannot be constructed from the paths that have been found thus far. While the algorithm can still use the paths to restrict the ends that the candidate template can take, the algorithm does not have knowledge about which edges should be included in the middle of the explanation. Thus, the algorithm must consider all combinations of edges from the schema to bridge these paths.

Bridging paths is beneficial because it can greatly reduce the space of candidate templates to test. In general, since the algorithm's performance is proportional to the number of candidates that must be tested, bridging improves performance because the start and end attribute constraints are pushed down in the algorithm. However, if only short paths are mined, but long explanation templates are desired (i.e., $n > 2\ell$), then the number of candidates exponentially increases with the length. Thus, for some length $n$, it is then no longer beneficial to bridge paths.

## 4. DEALING WITH MISSING DATA

So far we have only considered explanations that can be expressed solely in terms of the data stored in the database. Unfortunately, real databases are typically not perfectly curated. Information may be missing from the database, or relationships may not be recorded. For example, consider a nurse in a hospital who works directly with a doctor. When a patient has an appointment with the doctor, the appointment is recorded in the database, and we can use explanations of the type described in Section 2 to explain the doctor's accesses. Unfortunately, appointments are typically only scheduled with the doctor, not with the nurse. Thus, we cannot explain why the nurse accessed the patient's record, even though the access is appropriate.

To explain these types of accesses, we must deal with "missing" data in the database. One common type of missing data are the relationships between users of the database. While a database may store information such as the department each user works in, we show in Section 5.3.1 that additional information is still needed to explain accesses. Moreover, as we found in our data set, the nurse and doctor are assigned different department codes based on their job title. We hypothesize that information used to explain an access such as an appointment often is stored in the database with a reference to a single user, but that information can be used to explain why many other users access the data. Thus, if the database stored relationships among users, additional accesses could be explained.

By adding this missing data, the algorithms may introduce false positive explanations. In Section 5.3.2 we study the precision and recall trade-offs from adding missing data and show that even when missing data is added, the rate of false positive explanations is low due to the structure of the explanations.

A natural method to determine relationships between users of a database is to analyze user access patterns [10, 18]. In general, users who work together often access the same data. Using the log of accesses, we can automatically discover *collaborative groups* of users who access the same data often and use these groups to explain more accesses. For example, an explanation for why the nurse accessed the patient's medical record could be described as follows: the nurse accessed the patient's medical record because the nurse works with the doctor and the doctor had an appointment with the patient.

Next, we outline one possible approach to construct collaborative groups that we found to be effective for our data set. However, we note that there has been extensive work on clustering [13, 27], and alternative approaches are possible. In general though, we treat these algorithms as a black box that produces a set of relationships between users of the database. Once this data is plugged into the database, our explanation mining algorithms can incorporate the information to find additional supported templates.

## 4.1 Extracting Collaborative Groups

Given an access log, we can model the relationships between database users using a graphical structure. We use a method similar to that presented by Chen et al. [10]. Let a node in the graph represent a user. An edge exists between two users if the users access the same data. We assign weights to the edges to signify the strength of the users' relationship. To do this for a log of accesses that occur between some start and end time that has $m$ patients and $n$ users, we construct an $m \times n$ matrix $A$. The index $A[i, j]$ represents the inverse of the number of users (including user $j$) that accessed patient $i$'s record. More formally, if user $j$ does not access $i$'s record, then $A[i, j] = 0$, else:

$$A[i,j] = \frac{1}{\#\ users\ who\ accessed\ patient\ i's\ record}$$

The weight of an edge between user $u_1$ and user $u_2$ can be found in $W[u_1, u_2]$ where $W = A^T A$. Intuitively, $W[u_1, u_2]$ represents the similarity of two users' access patterns, relative to how often a particular record is accessed. Our current approach does not adjust the weight depending on the number of times a user accesses a specific record, but rather it only considers if a user accesses the record. A node's weight is defined as the sum of the connected edges' weights.

Given the graph structure, we can directly apply weighted graph clustering algorithms. Specifically, we use an algorithm that attempts to maximize the graph modularity measure [21]. Intuitively, optimizing for the graph modularity measure attempts to maximize the connections (and weights) for nodes within a cluster and minimize the connections between nodes that reside in different clusters. The algorithm is also parameter-free in the sense that it selects the number of clusters automatically.

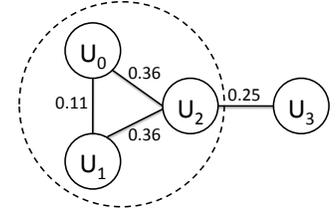

| Patient | User Ids |
|---------|----------|
| A | 0, 1, 2 |
| B | 0, 2 |
| C | 1, 2 |
| D | 2, 3 |

**Figure 5: (i) Example log of accesses (per patient), and (ii) the associated graphical representation with clustering.**

After running the clustering algorithm once, the algorithm outputs a set of clusters and an assignment of users to clusters. We can recursively apply the clustering algorithm on each cluster to produce a hierarchical clustering. Intuitively, clusters produced at the lower levels of the hierarchy will be more connected than clusters produced at higher levels. In Section 5.3.2 we show how this affects the precision and recall of explanations.

EXAMPLE 4.1. *Consider the log of accesses in Figure 5 that lists which users have accessed which patient's medical records. From the log, we can construct the matrix A. For example, $A[patient\ A, user\ 0] = \frac{1}{3}$ since three users accessed patient A's record. After evaluating $W = A^T A$, we find the edge weights that are labeled on the graphical representation. After running the clustering algorithm, users 0, 1 and 2 are assigned to the same cluster.*

After clustering, the table *Groups(Group_Depth, Group_id, User)* is added to the database. By applying a self-join on this table, the mining algorithms can use these groups to explain additional accesses.

EXAMPLE 4.2. *Nurse Nick's access of Alice's record in Figure 1 occurred because Nick works with Dr. Dave, and Dr. Dave had an appointment with Alice. The corresponding explanation template is expressed as follows:*

```
SELECT L.Lid,L.Patient,L.User,A.Date,G1.User
FROM Log L, Appointments A,
   Groups G1, Groups G2
WHERE L.Patient = A.Patient
   AND A.Doctor = G1.User
   AND G1.Group_id = G2.Group_id
   AND G2.User = L.User
```

## 5. EXPERIMENTAL EVALUATION

To test our ideas, we conducted an extensive experimental study using a real access log and database from the CareWeb system at the University of Michigan Health System. Our experiments aim to answer the following questions:

- *Do explanations (as we described them in Section 2.1) exist in real databases?* We find that explanations like those described in Example 2.1 occur in the real hospital database and can explain over 94% of the accesses in the log.

- *What missing information can be added to the database? Is this information useful for producing explanations?* Using the algorithm described in Section 4.1, we were able to find real-life collaborative groups, including the Michigan Cancer Center and the Psychiatric Services offices. After extending the database to include these collaborative groups, we were able to explain many more accesses.

- *Can we mine explanation templates efficiently?* We measure the performance of the one-way, two-way and bridged algorithms from Section 3 and find they are able to discover explanation templates automatically and efficiently. Moreover, the bridging optimization can improve performance in particular cases.

- *How effective are the mined explanation templates at correctly classifying future accesses?* We measure the precision and recall of the mined explanations and find that shorter explanations provide the best precision, but moderate recall. Longer explanations, including those that use the group information, can be used to improve recall.

- *Is the same set of explanation templates mined over time?* We find that the set of explanation templates discovered by the mining algorithms is relatively stable across time.

## 5.1 Implementation & Environment

Our system is a Python layer on top of PostgreSQL[4]. This layer constructs paths from the schema and executes queries on the database to determine an explanation's support. Clustering was performed with a Java implementation of the graph modularity algorithm. The experiments were executed on a dual core CPU with 4 GB of RAM, running Red Hat Linux.

## 5.2 Data Overview

We conducted an extensive experimental study using data from the University of Michigan Health System. To the best of our knowledge, ours is the first study to combine information from an access log with other information stored in a hospital's database for the purpose of explaining data accesses.

In our study, we used a de-identified log containing one week's worth of data accesses.[5] We also used de-idenfied data about the patients whose medical records were accessed during this period of time, including: (data set A) *Appointments*, *Visits*, *Documents*, and (data set B) *Labs*, *Medications* and *Radiology*. Information for these tables was extracted from the weeks around the time of the access.

The richness of the data set allows us to mine many different types of explanations. The log contained approximately 4.5M accesses (reads and writes), 124K distinct patients, and 12K distinct users. The number of distinct user-patient pairs is approximately 500K, which gives a user-patient density of $\frac{|user-patient\ pairs|}{|users| \times |patients|} = 0.0003$. To determine the reason for an access, we received data on approximately 51K appointments, 3K visits, 76K documents produced, 45K lab records, 242K medication records, and 17K radiology records. For example, the *Appointment* table contains a record for each appointment, including the patient, the doctor, and the date. The *Medications* table contains a record for each medication order, including the patient, the person who requested the medication, the person who signed for it, and the person who administered it. We were also given 291 descriptive codes describing which users worked in which departments such as *Pediatrics* and *Nursing-Pediatrics*.

---
[4] http://www.postgresql.org
[5] In all cases, protected health information (PHI) was removed, and patient and user IDs were coded in a way that did not permit the study team to link the data back to specific individuals.

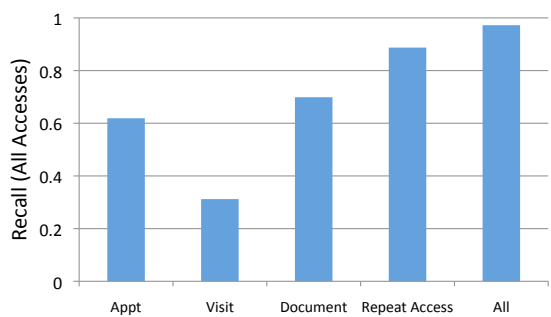

**Figure 6: Frequency of events in the database for all accesses.**

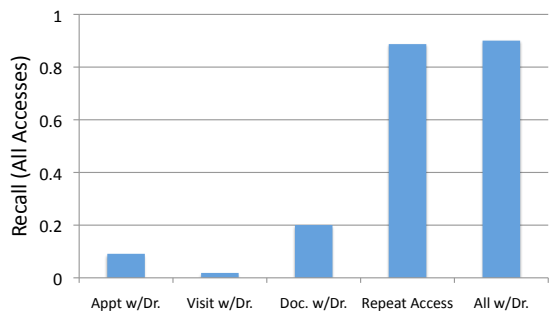

**Figure 7: Hand-crafted explanations' recall for all accesses.**

When we started the study, we initially requested the *Appointments*, *Visits*, and *Documents* tables, in addition to the log. However, after some preliminary analysis, we discovered that a large proportion of the unexplained accesses were by users who worked in departments that provide *consultation* services throughout the hospital (e.g., radiology, pathology, and pharmacy). Users in these departments often do not have appointments with patients. However, there is often an explicit request recorded in the database. Therefore, we expanded the study to also include the *Labs*, *Medications*, and *Radiology* tables, which maintain a record for each such request.

## 5.3 Results

### 5.3.1 Explanations in a Real Data Set

Our first set of experiments tests the fundamental hypothesis that accesses in the log can be explained using data stored elsewhere in the database.

We began by measuring the proportion of accesses in the log pertaining to a patient such that the patient had some type of $event$ recorded elsewhere in the database. In particular, we started by measuring the proportion of patients who had an appointment (Appt), visit, or document produced (Document). Figure 6 shows the frequency of these events in the log. (The recall of Appointment would be 1.0 if every patient whose record was accessed also had an appointment with someone listed in the database.) As expected, many patients had an appointment with someone or had a document produced (e.g., a doctor's note added to the file) by someone. Additionally, a majority of the accesses can be categorized as repeat accesses, meaning that the same user accessed the same patient's record for an additional time. When we combined all these events together, approximately 97% of all accesses corresponded to a patient who had some type of event in the database. Interestingly, a small percentage of the accesses did not correspond to a patient who experienced some type of event. We suspect that this is largely due to the incomplete data set. For example, appointments outside

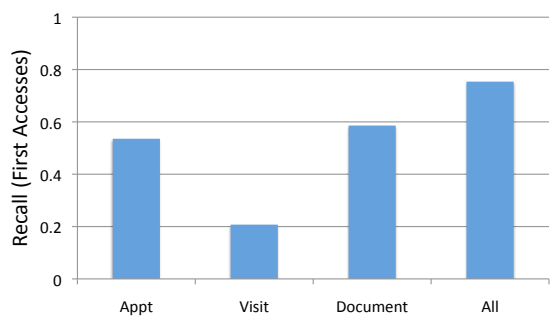

Figure 8: Frequency of events in the database for first accesses.

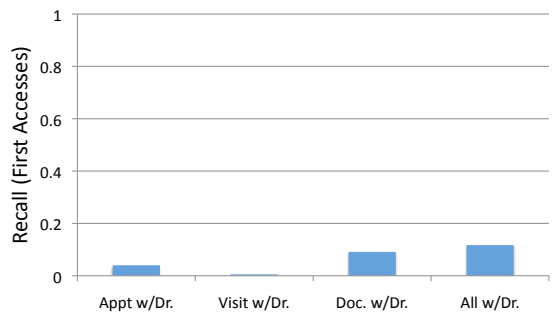

Figure 9: Hand-crafted explanations' recall for first accesses.

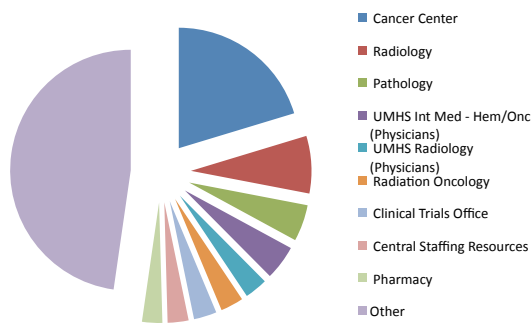

Figure 10: Collaborative Group I (Cancer Center)

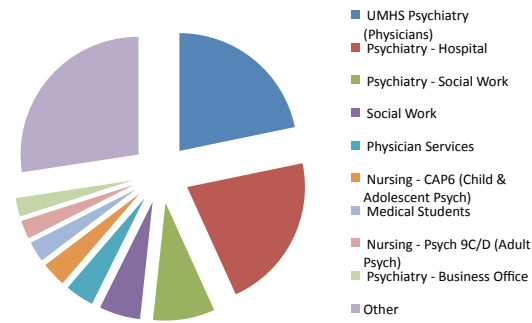

Figure 11: Collaborative Group II (Psychiatric Care)

of the study's timeframe were not considered.

Of course, these events do not constitute explanations since they do not necessarily connect the patient whose record was accessed to the specific user who accessed the record. (For example, a patient may have an appointment listed, but it may not be with the person who accessed her record.) To measure the proportion of accesses that can be explained using the approach described in Section 2.1, we hand-crafted a simple set of explanation templates, based on common reasons for medical records to be accessed, that test if the patient: (i) had an appointment with the specific doctor who accessed the record (Appt w/Dr.), (ii) had a visit with the doctor (Visit w/Dr.), (iii) had a document produced by the doctor (Doc. w/Dr.), or (iv) the access was a repeat access.

Figure 7 shows the recall for the explanations (i.e., proportion of the log records explained). While the repeat accesses can still explain a majority of the accesses, the recall of the other explanations is lower. This result is expected because the appointments, visits and documents produced typically only reference the primary doctor in charge of the patient's care. Therefore, using these basic explanation templates, we cannot explain why a nurse accesses a medical record. Even with this lower recall, these explanation templates can still explain 90% of the accesses.

Although repeat accesses make up a majority of the log, it is more challenging and interesting to explain why a user accesses a record for the first time. To do this, we analyzed all of the *first accesses* in the log, where a user accesses a patient's medical record for the first time. (Notice that since we only have a subset of the log, some accesses that are actually repeat accesses appear to be first accesses due to truncation.)

Figures 8 and 9 show the recall for the events and explanation templates among only the first accesses. When combined, the explanation templates for appointments, visits, and documents produced explain approximately 11% of first accesses (see the *All w/Dr.* bar in the chart). Ideally, we should be able to explain approximately 75% of the first accesses because 75% of the patients have some corresponding event in the database (see Figure 8). For the remaining 25% of the patients, we have no corresponding event. We attribute this result in large part to the incomplete data set.

In the next sections, we will show that it is possible to improve recall by adding missing data, and also by mining additional explanation templates.

### 5.3.2 Dealing With Missing Data

When a patient has an appointment, the appointment is scheduled with the doctor. However, the nurses who work with the doctor also typically access the patient's medical record. For this reason, we could only explain 11% of the first accesses, even though 75% of these patients have some associated event (e.g., an appointment with someone). To improve recall, we applied the algorithm described in Section 4.1 to cluster users who access similar medical records using the first six days of accesses in the log, and we added these collaborative groups to the database.

Since we were working with de-identified data, it was impossible to systematically verify the correctness of the resulting groups. However, a manual inspection of the groups suggests that the process was successful. For example, we studied the department codes present from the users in each group. Figures 10 and 11 show the department codes present in two of the 33 top-level groups. The first group clearly contains users who work in the Cancer Center, and the second group contains users who work in psychiatric care.

Interestingly, department codes themselves do not directly coincide with collaborative groups. For example, the *Medical Students* department code appears in the psychiatric care collaborative group. This makes sense because certain medical students were rotating through psychiatric care during the week when our log was collected, and they accessed the associated patients' medical records. However, medical students change rotations on a regular basis. This indicates that it would be incorrect to consider all medical students as their own collaborative group. It also indicates that we must update the collaborative groups from time to time in order

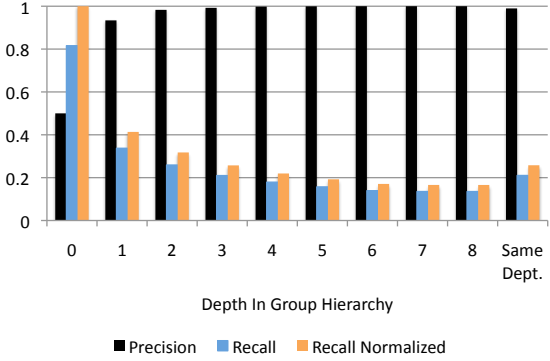

Figure 12: Group predictive power for first accesses (Data set A). Collaborative groups were trained using the first 6 days of the log; precision and recall were tested using the seventh day.

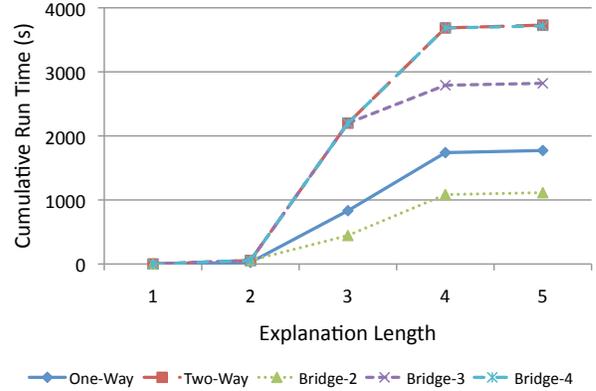

Figure 13: Mining performance (Data sets A & B, log days 1-6, T = 3, s = 1%)

to capture dynamic collaboration patterns.

Our goal in extracting collaborative groups is to improve explanation recall (i.e., the number of accesses that can be explained). As a baseline, we could assign all users to a single group; doing this, we are able to explain the 75% of first accesses where the patient has an event (see Figure 8). However, this approach has the consequence of potentially providing spurious *false positive* explanations if two users are not actually part of a collaborative group.

To measure the tradeoff between adding collaborative groups to improve recall and introducing false positives, we performed a simple experiment. We constructed a *fake* log that contains the same number of accesses as the real log. We generated each access in the fake log by selecting a user and a patient uniformly at random from the set of users and patients in the database. (Because the user-patient density in the log is so low, it is unlikely that we will generate many fake accesses that "look" real.) We then combined the real and fake logs, and evaluated the explanation templates on the combined log.

We define *recall* to be the proportion of real accesses returned by an explanation template from the set of all real accesses ($Recall = \frac{|Real\ Accesses\ Explained|}{|Real\ Log|}$). We define *precision* to be the proportion of real accesses that are in the set of all accesses returned ($Precision = \frac{|Real\ Accesses\ Explained|}{|Real+Fake\ Accesses\ Explained|}$). The *normalized recall* is the proportion of real accesses returned by an explanation template from the set of accesses we have information on ($Normalized\ Recall = \frac{|Real\ Accesses\ Explained|}{|Real\ Accesses\ With\ Events|}$). The normalized recall takes into account the fact we have a partial data set. In an ideal world, our explanation templates would observe precision and recall values close to 1.0.

We ended up with an 8-level hierarchy of collaborative groups, and we created the Groups table as described in Section 4.1. Using hand-crafted explanation templates that incorporate the groups (e.g., Example 4.2), we measured the precision, recall and normalized recall. Figure 12 shows the results for the groups at different levels of the hierarchy, measured using the first accesses from the seventh day of the log. Depth 0 refers to the naive approach of placing every user in a single group. Additionally, we included the hand-crafted explanation template that captures the idea that a user accesses a medical record because another user with the same department code has an appointment, visit or produced a document with the patient (e.g., explanation (B) from Example 2.1).

As expected, the top-level groups in the hierarchy (depths 0 and 1) result in higher recall, but lower precision. On the seventh day, the depth 0 group explains 81% of the first accesses. We also found that explanations based on collaborative groups outperformed explanations based on department codes, because users from different departments (e.g., *Pediatrics and Nursing-Pediatrics*) often work together.

In practice, depth 1 collaborative groups appear to strike a reasonable balance of high precision (>90%) and improved recall. For day seven in the log, if we consider explanations based on appointments, visits, documents produced, and repeat accesses (e.g., Figure 7), and we also include collaborative groups at depth 1, we are able to explain over 94% of all accesses.

### 5.3.3  Mining Explanations

Our next set of experiments measured the performance of the mining algorithms presented in Section 3. We ran the algorithms on the first accesses from the first six days of the log, with the combined data sets A and B, and the added group information. Based on an initial study, we set the support threshold to 1%. (A support threshold of 1% was sufficient to produce all of the explanation templates that we constructed by hand except one template where a doctor has a visit with a patient, which had a very small support.) We restricted the size of templates to $T = 3$ tables. We allowed self-joins on the Groups.Group_id attribute and the department code attribute. The algorithms utilized the optimizations described in Section 3.2.1. Due to how the data was extracted, data set B identifies users with a key *audit_id*, and data set A identifies users with a *caregiver_id*. We used a *mapping table* to switch from one identifier to the other. Thus, to deal with the slight difference in how the data was extracted, we did not count this added mapping table against the number of tables used.

We evaluated the algorithms based on their performance and their ability to find the hand-crafted explanation templates we previously constructed. Figure 13 shows the cumulative run time of the various algorithms by path length (the $length$ corresponds to the number of joins in the path). The algorithms mined explanations up to a length of five when the path included a self-join and the mapping table. Bridge-$\ell$ indicates that we used paths up to length $\ell$ for bridging. For our experimental setup, the Bridge-2 algorithm was the most efficient because it pushes the start and end constraints down in the algorithm. The one-way algorithm was faster than the two-way algorithm because the two-way algorithm considers more initial edges. Without the optimizations described in Section 3.2.1, the run time increases by many hours.

Each algorithm produced the same set of explanation templates. Moreover, it is worth noting that our mining algorithms were able to discover all the supported hand-crafted explanation templates we described in the paper such as appointments with doctors, appointments with users that work in the same department, and appoint-ments with users that are in the same group.

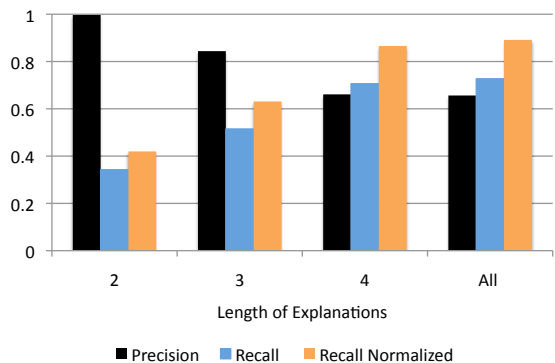

Figure 14: Mined explanations predictive power for first accesses (Data sets A & B, trained on days 1-6, tested on day 7)

It is important to note that this paper is not intended to be a full performance study. Rather, we intend this as a proof of concept, demonstrating that explanation templates can be mined automatically from a real data set. Therefore, the administrator's time can be saved if algorithms can find these explanation templates.

### 5.3.4 Predictive Power Of Explanations

Using the explanation templates mined from the first six days of accesses from Section 5.3.3 on data sets A and B, we tested the predictive power of the explanation templates on the seventh day of accesses using the same fake log that was described in Section 5.3.2. The goal is to determine if the mined explanation templates can correctly explain real accesses, while not spuriously explaining fake accesses. Figure 14 shows the results for explanations of various lengths, and the results when all of the explanations are tested together (All) for first accesses.

Explanation templates of length two have the best precision, while the recall is approximately 34% (42% normalized). These short explanations are like explanation (A) from Example 2.1, where the doctor has an appointment with the patient. The precision is high for these explanations because it is very unlikely that, for example, a fake access corresponds to an actual appointment. We believe this is a result of the user-patient density being so small.

It is also important to point out that the recall for these length-two explanation templates is higher when data set A and B are combined compared to when only data set A is included in Figure 12 (the recall increases from 13% to 34% when analyzing the first accesses for day seven). This change in recall shows that as more data is added to the database, we can explain additional accesses. With a complete data set, we argue that we can explain more accesses.

As explanation template paths get longer, the recall increases while the precision drops. Explanations of length three, which typically combine event information of two types (e.g., appointments and medications) have a recall of 51% (65% normalized). Explanation templates of length four, which use group information, increase the recall to 73% (89% when normalized). The precision drops since it is more likely that the user from a fake access corresponds, for example, to an appointment with another user that is in the same group. When all the explanations are combined and tested together, we find the recall and precision only change slightly from the length-four explanation templates because the longer explanations typically are more general versions of the shorter explanations. Therefore, the longer explanation templates explain most of the accesses that the shorter templates explain. For example, template (B) from Example 2.1 explains all those accesses explained by template (A).

|        | # Explanation Templates |       |       |       |                  |
|--------|-------------------------|-------|-------|-------|------------------|
| Length | Days 1-6                | Day 1 | Day 3 | Day 7 | Common Templates |
| 2      | 11                      | 11    | 11    | 12    | 11               |
| 3      | 241                     | 257   | 231   | 268   | 217              |
| 4      | 25                      | 25    | 25    | 27    | 25               |

Table 1: Number of explanations mined

We analyzed the department codes for which we could not explain the largest number of accesses. The top four departments were: Nursing-Vascular Access Service, Anesthesiology, Health Information Management, and Paging & Information Services. The users in the vascular access service department typically assist with IVs. Therefore, since our data set does not explicitly record why each nurse treated a patient and these nurses assist many different departments, it makes sense that the mined explanation templates could not explain their accesses.

For the evaluation, we used group information from any depth in the hierarchy. However, we observe that not every event type should use the same depth. For example, when only data set A was used, we had a precision of approximately 93% for depth 1, however when data set B was included, the precision dropped to 66%. Therefore, group information at one depth may be sufficient to explain an access with an appointment, but group information at another depth may be necessary to explain accesses with medication information to attain a desired level of precision. In the future, we will consider how to mine decorated explanation templates that restrict the groups that can be used to better control precision.

### 5.3.5 Stability of Explanations

Lastly, we measured the stability of the explanation templates that were mined over different time periods to determine if there exists a set of consistently occurring explanation templates. To do this, we ran the mining algorithms on different subsets of the log: days 1-6, day 1, day 3 and day 7. Table 1 shows the number of explanation templates produced per time period. For our data sets, the number of explanations that are produced is small enough for an administrator to manually analyze and approve those semantically correct explanations. Moreover, there is a set of *common explanation templates* that occur in every time period. Therefore, we believe the explanation templates mined represent generic reasons why medical records are accessed.

We did observe a small difference in the explanations mined across time periods. For example, on the seventh day, a twelfth length-two explanation template was added because there were more accesses corresponding to visits. We found larger variability in length-three explanations. This variation occurred from those explanation templates that connected two event types. For example, the path through radiology information to medication information occurred frequently on some days, but did not occur frequently during others.

## 6. RELATED WORK

Throughout this paper, we have focused on application-level access logs of the type generated, for example, by electronic health records systems. Typically, these logs store $(User, Patient)$ pairs, so it is easy to determine which accesses referenced a particular patient's record. A recent body of work has focused on the related problem of DBMS or SQL log auditing. In this case, logs are collected at the level of the DBMS recording the text of all SQL queries and updates; most commercial DBMSs now support this form of logging [17, 20, 26]. In contrast to application-level auditing, it is non-trivial to determine which logged queries accessed particular portions of the database, or particular records, and used these records in non-trivial ways. Various models and systems have been proposed to address this problem for audit logs collected at the level of the DBMS [4, 12, 15, 19].

Anomaly-based intrusion and misuse detection have been studied extensively in the past [8]. In the electronic health records domain, Chen et al. study how to detect anomalous insiders by analyzing an access log in a real hospital system [10]; they detect anomalous users by measuring the deviation of each user's access pattern from other users that access similar medical records. This work considers the user to be the unit of suspiciousness, deciding whether or not a user is behaving in an unexpected way. In contrast, we consider individual accesses, and we try to explain why each access occurs. This approach is more appropriate if, for example, hospital employees are generally well-behaved, but in some isolated cases they inappropriately access information (e.g., the Britney Spears [22] and Barack Obama cases [16]). In addition, our approach has the advantage of producing easily interpretable results, while the output of the anomaly detection mechanism does not clearly explain why a user is suspicious.

Auditing and misuse detection are often considered complementary to access control, which can be used to limit individual users' access to data [1, 23, 24]. Unfortunately, in health care, it is often infeasible to specify and enforce comprehensive access control policies due to the work environment. Instead, hospitals often maintain a log of accesses to detect misuse after the fact.

Recent work by Malin et al. propose learning access control policies from EHR access logs [18]. They use a data-driven approach to determine relationships between users and departments in the hospital. Using these relationships, for example, the system can determine the probability with which a patient's medical record will be accessed by a surgeon after it is accessed by an emergency room physician.

The mining algorithms presented in Section 3 have some similarities to previous work on pattern mining [5, 6, 14, 25], although existing algorithms do not solve our problem directly. The main differences between our problem and classical frequent pattern mining are as follows: First, we are mining connected paths between a start and end attributes in the schema, where the classical problem mines item sets. Second, our metric for frequency (support) is determined by the number of accesses in the log that are explained by the template. Therefore, every path we consider must reference the log. Additionally, the structure of the patterns that are mined and where the data is stored differs from the classical problem. For instance, the templates represent logical expressions that data in the database must satisfy. In contrast, the classical problem learns relationships between actual values. Lastly, the data is stored across multiple tables in the database, rather than in a single file of transactions.

In some ways, explaining why an access occurred is related to the basic motivation for data provenance. While our explanations attempt to explain why a user accesses data, provenance aims to explain where data came from, or how and why it was produced [7, 9, 11]. While there may be some connection between the two lines of research, provenance techniques do not directly solve the problem of explaining accesses in a database.

## 7. CONCLUSION

In this paper, we presented *explanation-based auditing*. Many systems, including EHRs, collect access logs. While this information is sufficient to explain who has accessed a particular piece of data (e.g., a patient's medical record), it is not usually enough to explain *why*. To address this problem, we introduced a framework that generates explanations automatically. Our work is based on the fundamental observation that accesses in specific classes of databases occur for a reason, and the reason can be inferred from data in the database. Thus, we model an explanation as a path that connects the data accessed to the user who accessed it, by way of data elsewhere in the database. Producing explanations for a large database can be time consuming. Instead, we provided algorithms to automatically mine explanation templates from the data. We evaluated our system on a real log and data set from the University of Michigan Health System. Using our model, we can explain over 94% of the accesses.

## 8. REFERENCES

[1] Oracle virtual private database (VPD). http://www.oracle.com.
[2] Personal communication with the University of Michigan Health System Compliance Office.
[3] Cerner Power Chart, 2011.
[4] R. Agrawal, R. J. Bayardo, C. Faloutsos, J. Kiernan, R. Rantzau, and R. Srikant. Auditing compliance with a Hippocratic database. In *VLDB*, pages 516–527, 2004.
[5] R. Agrawal and R. Srikant. Fast algorithms for mining association rules. In *VLDB*, pages 487–499, 1994.
[6] R. Agrawal and R. Srikant. Mining sequential patterns. In *ICDE*, pages 3–14, 1995.
[7] P. Buneman, S. Khanna, and W. chiew Tan. Why and where: A characterization of data provenance. In *ICDT*, pages 316–330, 2001.
[8] V. Chandola, A. Banerjee, and V. Kumar. Anomaly detection: A survey. *ACM Computing Surveys*, 41, 2009.
[9] A. Chapman and H. V. Jagadish. Why not? In *SIGMOD*, pages 523–534, 2009.
[10] Y. Chen and B. Malin. Detection of anomalous insiders in collaborative environments via relational analysis of access logs. CODASPY, pages 63–74, 2011.
[11] J. Cheney, L. Chiticariu, and W. C. Tan. Provenance in databases: Why, how, and where. *Foundations and Trends in Databases*, 2009.
[12] D. Fabbri, K. LeFevre, and Q. Zhu. Policyreplay: Misconfiguration-response queries for data breach reporting. In *VLDB*, pages 36–47, 2010.
[13] S. Fortunato. Community detection in graphs. *Physics Reports*, 486(3-5):75 – 174, 2010.
[14] J. Han, J. Pei, and Y. Yin. Mining frequent patterns without candidate generation. In *SIGMOD*, pages 1–12, 2000.
[15] R. Kaushik and R. Ramamurthy. Efficient auditing for complex SQL queries. In *SIGMOD*, pages 697–708, 2011.
[16] G. Kessler. Two fired for viewing Obama passport file. *Washington Post*, March 21 2008.
[17] D. Kiely. SQL Server 2008 Security Overview for Database Administrators. SQL Server Technical Article, October 2007.
[18] B. Malin, S. Nyemba, and J. Paulett. Learning relational policies from electronic health record access logs. *Journal of Biomedical Informatics*, 44(2):333–342, 2011.
[19] R. Motwani, S. Nabar, and D. Thomas. Auditing SQL queries. In *ICDE*, pages 287–296, 2008.
[20] A. Nanda. Fine-grained auditing for real-world problems. *Oracle Magazine*, 2004.
[21] M. Newman. Analysis of weighted networks. *Phys. Rev. E*, 70, 2004.
[22] C. Ornstein. Hospital to punish snooping on Spears. *Los Angeles Times*, 2008.
[23] S. Rizvi, A. Mendelzon, S. Sudershan, and P. Roy. Extending query rewriting techniques for fine-grained access control. In *SIGMOD*, pages 551–562, 2004.
[24] R. S. Sandhu, E. J. Coyne, H. L. Feinstein, and C. E. Youman. Role based access control models. *IEEE Computer*, 29:38–47, 1996.
[25] S. Sarawagi, S. Thomas, and R. Agrawal. Integrating association rule mining with relational database systems: alternatives and implications. In *SIGMOD*, pages 343–354, 1998.
[26] T. J. Wasserman. DB2 UDB Security Part 5: Understanding the DB2 Audit Facility, March 2006.
[27] R. Xu and D. Wunsch II. Survey of clustering algorithms. *IEEE Transactions on Neural Networks*, 16:645–678, 2005.